\documentclass[twocolumn,prb,preprintnumbers,showpacs,amsmath,amssymb]{revtex4}

\usepackage{graphicx}
\usepackage{dcolumn}
\usepackage{bm}

\begin{document}

\title{Hardness of T-carbon: Density functional theory calculations}

\author{Xing-Qiu Chen$^1$}
\email[Corresponding author: ]{xingqiu.chen@imr.ac.cn},
\author{Haiyang Niu$^1$}
\author{Cesare Franchini$^{1,2}$}
\author{Dianzhong Li$^1$}
\author{Yiyi Li$^1$}

\affiliation{$^1$ Shenyang National Laboratory for Materials
Science, Institute of Metal Research, Chinese Academy of Sciences,
Shenyang 110016, P. R. China}

\affiliation{$^2$University of Vienna, Faculty of Physics and Center
for Computational Materials Science, Vienna, Austria}

\date{\today}

\begin{abstract}
We revisit and interpret the mechanical properties of the recently
proposed allotrope of carbon, T-carbon [Sheng \emph{et al.}, Phys.
Rev. Lett., \textbf{106}, 155703 (2011)], using density functional
theory in combination with different empirical hardness models. In
contrast with the early estimation based on the Gao's model, which
attributes to T-carbon an high Vickers hardness of 61 GPa comparable
to that of superhard cubic boron nitride (\emph{c}-BN), we find that
T-carbon is not a superhard material, since its Vickers hardenss
does not exceed 10 GPa. Besides providing clear evidence for the
absence of superhardenss in T-carbon, we discuss the physical
reasons behind the failure of Gao's and \v{S}im$\rm\mathring{u}$nek
and Vack\'a\v{r}'s (SV) models in predicting the hardness of
T-carbon, residing on their improper treatment of the highly
anisotropic distribution of quasi-\emph{sp}$^3$-like C-C hybrids. A
possible remedy to the Gao and SV models based on the concept of
superatom is suggest, which indeed yields a Vickers hardness of
about 8 GPa.

\end{abstract}

\pacs{64.60.My, 64.70.K-, 62.25.-g, 62.20.Qp}

\maketitle

\begin{figure}[hbt]
\begin{center}
\includegraphics[width=0.45\textwidth]{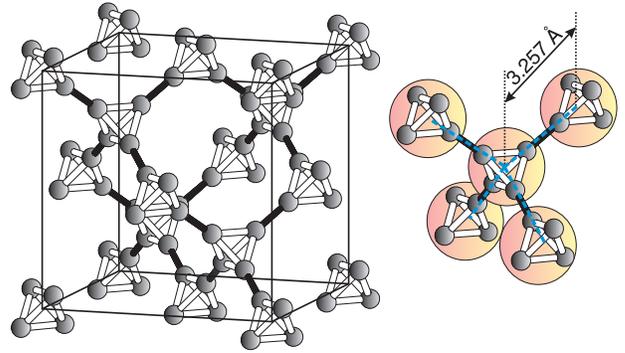}
\end{center}
\caption{Lattice structure of T-carbon (space group
Fd$\overline{3}$m). By considering each carbon tetrahedron (C4 unit)
as an artificial superatom, the corresponding structure is isotypic to
that of diamond. The local environment of each superatom is
illustrated in the right panel.} \label{fig:0}
\end{figure}

Recently, on the basis of first-principles calculations Sheng
\emph{et al.} proposed a carbon allotrope which they named T-carbon
\cite{Sheng}. Strictly speaking, its actual stability needs a highly
large negative pressure which is far beyond currently available
technologies. Structurally, this phase can be obtained by
substituting each carbon atom in diamond with a carbon tetrahedron
(Fig. \ref{fig:0}), and thus crystallizes in the same cubic
structure of diamond (space group Fd$\overline{3}$m)) with the
carbon atoms at the Wyckoff site 32\emph{e} (0.0706, 0.0706,
0.0706). It has been noted that T-carbon has a large lattice
constant of 7.52 \AA\ and a low bulk modulus of \emph{B} =169 GPa,
only 36.4\% of the bulk modulus of diamond\cite{Sheng}. In
particular, its equilibrium density, 1.50 g/cm$^3$, is the smallest
among diamond (cubic and hexagonal diamond) \cite{Sheng}, graphite
\cite{Sheng}, \emph{M}-carbon \cite{Li2}, bct-C$_4$ \cite{333},
\emph{W}-carbon \cite{444}, chiral-carbon \cite{Pickard} as well as
the newly proposed dense \emph{hp}3-, \emph{tI}12- and
\emph{tP}12-carbon \cite{Zhuqiang} phases. This results in an highly
porous structural pattern, which can be viewed as a diamond-like
array of superatoms (tetrahedral C4 clusters), as depicted in Fig.
\ref{fig:0}. Given this peculiar clusterized arrangement of atoms
exhibiting a quite low shear modulus of \emph{G} = 70 GPa
\cite{Sheng}, it is very surprising that T-carbon was predicted to
be superhard, with an exceptionally high Vickers hardness (H$_v$) of
61.1 GPa\cite{Sheng}, comparable to that of superhard cubic boron
nitride (\emph{c}-BN).

The aim of our present study is to elucidate the origin of this
anomalous hardness. We do this by exploring in details the
mechanical properties of T-carbon through the application of several
different empirical approaches: the Gao's formula\cite{555}, the SV
model\cite{666} and our recently proposed empirical treatment based
on the Pugh's modulus ratio\cite{ChenHv}. Our systematic analysis
provides an unambiguous and physically sound results: T-carbon is
not hard. We will show that the conventional application of Gao and
SV models leads to a much too high Vickers hardness, H$_v^{\rm
Gao}$=61.1 GPa and H$_v^{\rm SV}$=40.5 GPa, substantially
overestimated with respect to the value obtained using our
formalism, H$_v^{\rm Chen}$=5.6 GPa. The prediction of a low Vickers
hardness in T-carbon is consistent with the estimation of a low
shear strength (7.3 GPa along the (100)$<$001$>$ slip system), which
represents the upper bound of the mechanical strength.

The improper assignments derived by a conventional application of
the Gao and SV models can be attributed to the fact that these two
models assume that the chemical bonds, which are significant for
hardness, are distributed {\em uniformly} in the lattice. But
in T-carbon, as already pointed out be Sheng {\em et
al.}\cite{Sheng}, though the carbons atoms are tetrahedrally
coordinated and apparently resembling a three-dimensional
quasi-\emph{sp}$^3$-like hybrid \cite{Sheng}, their bonds are
ordered in an extremely anisotropic and porous framework, highly
different from the bonding distribution in ideal
\emph{sp}$^3$-hybrid. We propose a remedy to cure the limitations
of Gao and SV models in dealing with anisotropic and porous systems
by assuming each carbon tetrahedron cluster as an artificial
superatom. Indeed, this cluster-like approach leads to low Vickers
hardness in the range of 7-8 GPa, in agreement with the estimated
value of 5.6 GPa using our proposed model \cite{ChenHv}.

All calculations were performed using the Vienna \emph{ab initio}
Simulation Package (VASP)~\cite{G1} in the framework of density
functional theory (DFT), and  we adopted the Perdew, Burke and
Ernzerhof approximation\cite{G2} to treat the exchange-correlation
kernel. Well converged results were obtained using an energy cut-off
of 500 eV and a k-point grid 11$\times$11$\times$11\cite{cf}. The
DFT results were then employed as input for the three different
hardness empirical models, with which we have computed the Vickers
hardness H$_v$:

(a) Gao's model\cite{555}:
\begin{equation}
H_v^{\rm Gao} = 350[(N_e^{2/3})e^{-1.191f_i}/(d^{2.5})]
\label{eq:gao}
\end{equation}
where $N_e$ is the electron density of valence electrons per
\AA\,$^3$, $d$ is the bond length and $f_i$ is the ionicity of the
chemical bond in a crystal scaled by Phillips. As already mentioned,
this model gives H$_v^{Gao}$=61.1 GPa (Ref. \onlinecite{Sheng}).

(b) SV model\cite{666}:
\begin{equation}
H_v^{\rm SV} = \frac{C}{\Omega}\sqrt{e_ie_i}/(d_{ii}n_{ii})
\label{eq:sv}
\end{equation}
where $C$ is the constant of 1550 and $\Omega$ is the equilibrium
volume of T-carbon. $e_i$ = $Z_i$/$R_i$ represents the reference
energy, with $Z_i$ indicating the valence number of element $i$. For
carbon $e_i$ = 4.121 (taken from Ref. \onlinecite{666}). $n_{ii}$
and $d_{ii}$ are the number of bonds and bonding lengths between
atom $i$. In T-carbon, each carbon has four nearest-neighbors with
two different bonding lengths: three intratetrahedron carbon-carbon
bonding length of 1.502 \AA\, and one intertetrahedron bonding
length of 1.417 \AA\, \cite{Sheng}. By using the average bonding
length of 1.48075 \AA\ we obtained $H_v$ =
(1550/26.5785)$\times$4.121/(1.48075$\times$4) = 40.5 GPa, which is
33.5\% smaller than the corresponding Gao's value.\cite{Sheng}

\begin{figure}[hbt]
\begin{center}
\includegraphics[width=0.45\textwidth]{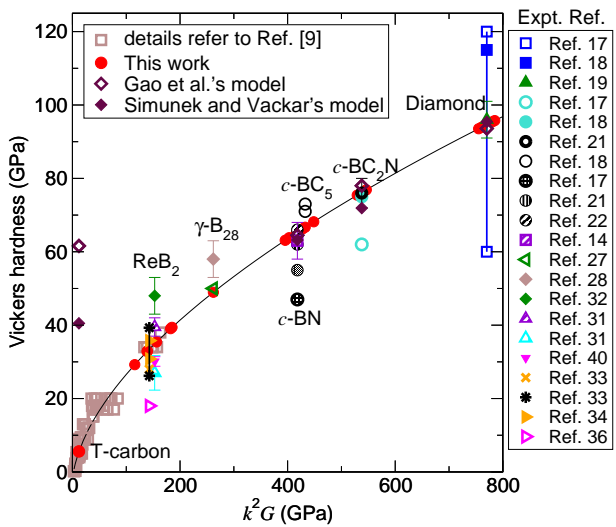}
\end{center}
\caption{Vickers hardness $H_v$ as a function of a product
($k$$^2$$G$) of the squared Pugh's modulus ratio (\emph{k} =
\emph{G}/\emph{B}) and shear modulus (\emph{G}). The curve
corresponds to the empirical relation of Eq. \ref{eq:chen} (For
other data and more details, see Ref. \onlinecite{ChenHv}). Elastic
moduli and experimental Vickers hardness are collected in Table
\ref{tab1}. Note the huge discrepancies among the three theoretical
estimations for T-carbon. } \label{fig:1}
\end{figure}

(c) Chen's model\cite{ChenHv}. This is the empirical formula which
we have recently proposed, based on the Pugh's modulus ratio
$k$=$G$/$B$\cite{888}:
\begin{equation}
H_v^{\rm Chen} = 2(k^2G)^{0.585}-3 \label{eq:chen}.
\end{equation}
This model not only reproduced well the experimental values of
Vickers hardness of a series of hard materials including all
experimentally verified superhard materials (see Fig. \ref{fig:1}
and Table \ref{tab1}), but also provides a theoretical foundation of
Teter's empirical correlation \cite{72} in its simplified form
\cite{ChenHv}.

Before discussing the results for T-carbon we start by presenting
some general considerations regarding the calculation of the Vickers
hardness and the trustability of our proposed model\cite{ChenHv}.
Hardness is a highly complex property, which depends on the
loading force and on the quality of samples (i.e., presence of
defects such as vacancies and dislocations). Because Vickers
hardness is experimentally measured as a function of the applied
loading forces, the saturated hardness value (or experimental
load-invariant indentation hardness) is usually considered to be the
hardness value of a given material. Therefore, the theoretically
estimated Vickers hardness within Gao's, SV's and Chen's models
should be directly compared to the experimentally saturated hardness
value of polycrystalline materials. An overview on the experimental
and theoretical values of H$_v$ for the experimentally verified
superhard materials (diamond, \emph{c}-BC$_2$N, \emph{c}-BN,
\emph{c}-BC$_5$, $\gamma$-B$_{28}$) is summarized in Fig.
\ref{fig:1} and Tab. \ref{tab1}. The experimental results are highly
scattered, reflecting the inherent difficulties in achieving a
trustable and precise estimation of hardness. For instance, the
reported values for the hardness of diamond, the archetype superhard
materials, range from 60 GPa to 120 GPa\cite{62,82,92,102}. Similar
trends have been observed for the other two well-known superhard
materials \emph{c}-BC$_2$N and \emph{c}-BN. The most typical case is
probably ReB$_2$, whose actual hardness has been extensively debated
\cite{202,212,222,232,242,252,262,272,282,292,302,312,322,Chen08}
after the first value of its Vickers hardness (48 $\pm$ 5 GPa at the
loading force of 0.49 N) was reported \cite{222}. Depending on
different samples, synthetic methods and measurement technique, the
obtained values range from 18 to 48 GPa (Table \ref{tab1}). In
contrast to experiment, theoretical estimations of the Vickers
hardness given by different models \cite{555,666,ChenHv} agree
within few GPa, including the data obtained by our proposed model
(Eq. \ref{eq:chen}). Overall, the comparative trend displayed in
Fig. \ref{fig:1} provides robust evidence for the reliability of our
proposed formalism \cite{ChenHv}.

Now, let's turn the attention to T-carbon. By using the values of
the shear and bulk moduli from Ref.\onlinecite{Sheng} as input
(\emph{B} = 169 GPa and \emph{G} = 70 GPa) for Eq. \ref{eq:chen} we
obtain a Vickers hardness of 5.6 GPa, dramatically smaller than the
corresponding Gao (61.1 GPa) and SV (40.5 GPa) estimations.
Furthermore, we noted that Sneddon defined the concept of ideal
elastic hardness by \begin{math}
H_{id}=\frac{Ecot\phi}{2(1-v^2)}\end{math} where \emph{E} is Young's
modulus, $v$ is Poisson's ratio and cot$\phi$ $\approx$ 0.5 for the
standard pyramid indentation \cite{Hid} and suggested that the real
hardness would be (0.01$\sim$0.2)H$_{id}$ at high loads \cite{Hid}.
Utilizing this definition and the derived \emph{E} = 185 GPa, the
real hardness for T-carbon should be in the range from 0.5 GPa to 10
GPa, in agreement with our obtained value. In particular, it still
needs to note that the occurrence of this serious discrepancy among
the three different methods (Gao's, SV's and Chen's models), which
is not observed for the other test cases of Fig. \ref{fig:1} and
Table \ref{tab1}, urges for a clarification aiming to discern which
method provides the more reliable description of the hardness of
T-carbon and, consequently, to help us to answer a naturally arising
question: is T-carbon a real superhard material?

\begin{table}
\caption{Comparison between measured ($H_v^{Exp}$) and theoretically
computed values of the Vickers' values (in GPa), along with
available bulk modulus (B, GPa), shear modulus (G, GPa) and Pugh's
modulus ratio $k$= $G$/$B$. }
\begin{ruledtabular}
\begin{tabular}{cccccccccccccccccc}
 &  \emph{G} &  \emph{B} &  \emph{k} & H$_{v}^{\rm Chen}$
& $H_v^{\rm Exp}$ &
$H_v^{\rm Gao}$ & $H_v^{\rm SV}$ \\
\hline
Diamond & 536$^a$ & 442$^a$ & 1.211 &  95.7 & 60-120$^d$ & 93.6 & 95.4  \\
        & 548$^a$ & 466$^a$ & 1.178 &  93.9 & 115$^e$ & & \\
        & 520$^b$ & 432$^b$ & 1.205 &  93.5 & 95$\pm$5$^f$ & & \\
        & 535$^c$ & 443$^c$ & 1.208 &  95.4 & & & \\
\hline
$c$-BC$_2$N & 446$^g$ &  403$^g$ & 1.107 & 76.9 & 62$^d$,75$^e$ & 78 & 71.9 \\
            & 445$^c$ &  408$^c$ & 1.091 & 75.4 & 76$\pm$4$^{e,h}$ & &  \\
\hline
$c$-BC$_5$ & 394$^i$ & 376$^i$ & 1.048 &  66.7& 71$^e$,73$^e$ & &  \\
\hline
$c$-BN & 405$^j$ & 400$^j$ &  1.014 &  65.2 & 47$^d$  &  64.5  &  63.2  \\
       & 403$^a$ & 404$^a$ &  0.999 &  63.8 & 55$^h$ & & \\
       & 382$^a$ & 376$^a$ & 1.017  & 63.1 & 62$^h$ & & \\
       & 404$^k$ & 384$^k$ & 1.053  & 68.2 &66$^l$ & & \\
       & 409$^c$ & 400$^c$ & 1.023 &  66.2 & 63$\pm$5$^c$ & & \\
\hline
$\gamma$-B$_{28}$ & 236$^m$ &  224$^m$ &  1.054 &  49.0 & 50$^n$,58$\pm$5$^o$ & &  \\
\hline
ReB$_2$ & 273$^p$ & 382$^p$ &  0.715 &  32.9 & 48$\pm$5$^s$ & & \\
        & 273$^r$ & 383$^r$ &  0.712 &  32.8 & 39.5$\pm$2.5$^r$ & & \\
        & 183$^r$ & 230$^r$ &  0.795 &  29.3 & 27$\pm$4.7$^r$ & & \\
        & 289$^x$ & 365$^x$ &  0.794 &  39.0 & 37.2-40.5$^t$ & & \\
        & 283$^y$ & 264$^z$ &  0.808 &  39.4 & 28$^t$ & & \\
        & 350$^y$ & 343$^z$ &  0.769 &  35.4 & 39.3-26.2$^u$ & & \\
        &           &           &        &       & 30.8-35.8$^v$ & &  \\
        &           &           &        &       & 18$^w$ & &  \\
        &           &           &        &       & 30.1$\pm$1.3$^A$ & &  \\
        &           &           &        &       & 37$\pm$1.2$^B$ & &
        \\
\hline T-carbon     & 70$^C$    & 169$^C$&  0.414&   5.6 &  & 61.1 &
40.5
\\
\end{tabular}
\end{ruledtabular}
$^a$ Ref. \cite{52}, $^b$ Ref. \cite{62}, $^c$ Ref. \cite{72}, $^d$
Ref. \cite{82}, $^e$ Ref. \cite{92}, $^f$ Ref. \cite{102}, $^g$ Ref.
\cite{112}, $^h$ Ref. \cite{122}, $^i$ Ref. \cite{132}, $^j$ Ref.
\cite{142}, $^k$ Ref. \cite{152}, $^l$ Ref. \cite{162}, $^m$ Ref.
\cite{172}, $^n$ Ref. \cite{182}, $^o$ Ref. \cite{192}, $^p$ Ref.
\cite{202}, $^q$ Ref. \cite{212}, $^r$ Ref. \cite{222}, $^s$ Ref.
\cite{232}, $^t$ Ref. \cite{242}, $^u$ Ref. \cite{252}, $^v$ Ref.
\cite{262}, $^w$ Ref. \cite{272}, $^x$ Ref. \cite{282}, $^y$ Ref.
\cite{292}, $^z$ Ref. \cite{302}, $^A$ Ref. \cite{312}, $^B$ Ref.
\cite{322}, $^C$ Ref. \cite{Sheng}. \label{tab1}
\end{table}

\begin{figure}[hbt]
\begin{center}
\includegraphics[width=0.38\textwidth]{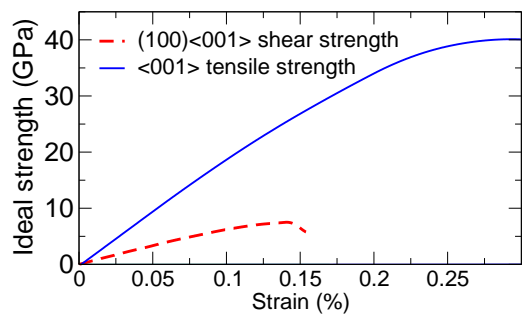}
\end{center}
\caption{DFT calculated ideal tensile and shear strengths of
T-carbon.} \label{fig:2}
\end{figure}

An useful concept for understanding strong mechanical strength --
but still relying on elastic properties -- is based on ideal shear
and tensile strengths \cite{Oganov}, at which a material is getting
unstable under direction-dependent deformation strains
\cite{Roundy}. To shed some light on the nature of T-carbon we have
thus investigated ideal tensile strength along the $<$001$>$
direction and shear strength along the (100)$<$001$>$ slip system.
We found that a tensile strength of 40.1 GPa along the $<$001$>$
direction and a shear strength of 7.3 GPa in the (100)$<$001$>$ slip
system (see Fig. \ref{fig:2}). Therefore, we can conclude that the
failure mode in T-carbon is dominated by the shear deformation type
in the (100)$<$001$>$ slip system. The calculated shear stress of
7.3 GPa basically sets the upper bound on its mechanical strength at
zero pressure \cite{Roundy,Oganov}, because the ideal strength is
the stress where a defect-free crystal becomes unstable and
undergoes spontaneous plastic deformation. It is well-known that the
measurement of hardness has to first encounter the elastic
deformation and then experience permanent plastic deformation.
Therefore, it can be conjectured that the hardness of T-carbon
should not exceed 7.3 GPa. These arguments provide a strong support
for our estimated Vickers hardness of 5.6 GPa on the basis of Eq.
\ref{eq:chen}.

In order to gain further insights on this intricate subject and to
reach a consistent and satisfactory conclusion on the hardness of
T-carbon we consider now the relation between hardness and
brittleness on the basis of the Pugh's modulus ratio \cite{888}.
There is no doubt that all experimentally verified superhard
materials, such as diamond, $c$-BN, $c$-BC$_2$N, $\gamma$-B$_{28}$
and $c$-BC$_5$ are intrinsically brittle. As shown in Table
\ref{tab1} the Pugh's modulus ratio \cite{888} of these superhard
materials ($k$ = 1.211-1.178 (diamond), 0.999-1.053 ($c$-BN),
1.107-1.091 ($c$-BC$_2$N), 1.054 ($\gamma$-B$_{28}$) and 1.048
($c$-BC$_{5}$)) are larger than 1.0. They clearly obey to the
empirical relation that considers the Pugh's modulus ratio as an
indicator of the brittleness or ductility of materials. The higher
\emph{k} the more brittle (and less ductile) the material is. Pugh
still proposed, when \emph{k} is larger than 0.571 the materials are
brittle and with \emph{k} being less than 0.571 the materials are
ductile \cite{888}. This relation has been extensively applied not
only to metals and alloys but also to high-strength materials. In
the case of T-carbon, the calculated Pugh's modulus ratio \emph{k} =
0.414 is smaller than 0.571, clearly in the range of ductility. The
ductile behavior of T-carbon is a further indication of its
non-superhardness.

On the basis of the above consideration we can now understand why
T-carbon is not an superhard material. One common feature of
superhard materials is that they not only need a three-dimensional
network composed of short, strong, and covalent bonds \cite{Kaner}
but also have a \emph{uniform} distribution of strong covalent
bonds. The prototypical example is diamond, which is characterized
by an isotropic array of tetrahedrally bonded $sp^3$ carbon atoms.
Conversely, in soft graphite the \emph{sp}$^2$-type covalent bonds,
though strong, are localized in two-dimensional sheets. At first
glance, T-carbon seems to be a good candidate for superhardness
since each carbon atom has four nearest-neighboring carbon
tetrahedrally bonded by short and strong carbon-carbon covalent
bonds. However, due to the extreme-anisotropic arrangement of these
carbon-carbon bonds and the associated formation of a large
proportion of porosity in lattice space as well as the low density
of bonds, the framework of T-carbon be more easily bendable in
comparison with that of diamond, as manifested by its low shear
strength.

Having this in mind, we can look back Gao's and SV models. Although
these two models perform very well for many hard materials, they
deliver questionable numbers for T-carbon in sharp contrast with our
findings, as we have documented above. The reason for this apparent
failure is that in these two models all bonds are treated as
\emph{uniformally} distributed in the lattice space. Clearly, this
constrain will not affect the predictions for isotropic material but
it will be inadequate to describe the hardness of
extremely-anisotropic compounds such as T-carbon. However, if
we give a closer look to each individual C4 tetrahedron unit (see
Fig. 1), the distribution of six strong carbon-carbon covalent bonds
within each C4 unit is highly dense. It is therefore trustfully
expected, that the Vickers hardness of each individual C4 unit can
be comparable (or even harder) to that of diamond because its bonds
density and strengths within each C4 unit are higher than those of
diamond. The strength and rigidity of each individual C4 unit appear
to be such strong that it cannot be broken easily. Based on this
fact, in order to render Gao's and SV's methods applicable to
T-carbon, each carbon tetrahedron (C4 unit) is considered to be an
artificial superatom (See right panel of Fig. \ref{fig:0}). The
cubic unit cell of T-carbon consists of eight superatoms and each
superatom has four nearest neighbors with the bonding length of $d$
= 3.257 \AA\,. In terms of Gao's and SV's methods, this
distance d should be the bonding length between exact atomic
positions with positively charge cores, representing the real force
center of each atom. Based on our assumption, the $d$ distance is
defined as the spatial separation between two nearest  neighbor
superatom positions, $d$=3.257 \AA. Although it remains disputable
whether the center of mass of the C4 superatom could be assigned its
real force center (thus allowing the applicability of Gao's and SV's
models) the high strength and rigidity of each individual C4 unit
manifested by the dense and strong carbon-carbon bonds seem to
validate this assumption of d distance. Obviously, each superatom
contains 16 valence electrons, and $N_e$ = 8/26.61 = 0.3. By
inserting these values of $d$ and $N_e$ in Gao's formula (Eq.
\ref{eq:gao}), we derive a Vickers hardness of 8.2 GPa, in agreement
with our value of 5.6 GPa. To apply the same adjustment to the SV
model one needs to define the crucial parameter $R_i$ for the
superatom. From our first-principles calculations, it can be
inferred that $R_i$ = 2.32 \AA\, represents the optimum radius
containing all 16 valence electrons for each superatom. By inserting
$e_i$ = 16/2.32 = 6.896 in Eq. \ref{eq:sv} a Vickers hardness of 7.7
GPa is obtained, again in agreement with our analysis. Within this
superatom approach, all three methods discussed in the present paper
convey the same answer: T-carbon is not superhard. The anomalous
behavior of Gao and SV models observed in Fig.\ref{fig:1} for
T-carbon is cured and the general agreement among the three Gao, SV
and Chen models is re-established. This provides clear evidence that
the hardness of T-carbon should not exceed 10 GPa.

{\bf Acknowledgements} We greatly appreciate the thought of the
treatment of ``superatom'' to successfully apply Gao et al.'s model
to hardness of T-carbon from Prof. Faming Gao and useful discussions
with Prof. Gang Su and Dr. D.-E. Jiang for his critical reading. X.
-Q. C. acknowledges the support from the ``Hundred Talents Project''
of CAS and the NSFC (Grant No. 51074151). The authors also
acknowledge the computational resources from the Supercomputing
Center (including its Shenyang Branch in the IMR) of CAS.

\end{document}